\documentclass[twocolumn,aps,prl,superscriptaddress,showpacs,floatfix,amsmath]{revtex4-2}
\usepackage{epsfig}
\usepackage{amsmath}
\usepackage{bm}

\usepackage{here}
\usepackage{float}
\usepackage{graphicx}
\usepackage[linkcolor=black,urlcolor=blue,citecolor=blue,colorlinks=true]{hyperref}
\usepackage{xcolor}

\usepackage{subfigure}
\usepackage{comment}
\usepackage{esvect}
\usepackage{gensymb}% for degree
\usepackage{multirow}
\usepackage{dcolumn}   % needed for some tables

\usepackage[normalem]{ulem}
\begin{document}
	
\title{Overdoping YBa$_2$Cu$_3$O$_7$ via a heterostructure with La$_{0.67}$Sr$_{0.33}$MnO$_3$}
	
\author{Ankita Singh, Sawani Datta, Ram Prakash Pandeya, Srinivas C. Kandukuri, Mahesh Gokhale, and Kalobaran Maiti}
\altaffiliation{Corresponding author: kbmaiti@tifr.res.in}
\affiliation{Department of Condensed Matter Physics and Materials Science, Tata Institute of Fundamental Research, Homi Bhabha Road, Colaba, Mumbai-400005, India.}
	
%\maketitle
	
\begin{abstract}
YBa$_2$Cu$_3$O$_x$, the first superconductor discovered with $T_c$ higher than 77 K, is among the most complex cuprates having both CuO chains and plains in the structure. YBa$_2$Cu$_3$O$_7$ (YBCO) exhibits slightly overdoped behavior and further doping is difficult as all the lattice sites in the CuO chains are occupied. We have grown high quality single crystalline films of YBCO and bilayer La$_{0.67}$Sr$_{0.33}$MnO$_3$ (LSMO)/YBCO exhibiting superconductivity in \emph{both} the cases. Photoemission spectra reveal different surface and bulk electronic structures; the difference reduces in the bilayer. Evidence of charge transfer across the bilayer interface is observed in the valence band and core level spectra indicating an \textit{overdoped} condition in YBCO. While superconductivity in the presence of magnetic order in the bilayer is puzzling, this pathway to reach overdoped regime in YBCO opens up a new landscape to probe the exotic physics of unconventional superconductivity.
\end{abstract}
	
\maketitle
	
%\section{Introduction}

\begin{figure}[h!]
 \vspace{-2ex}
\includegraphics [scale=0.45]{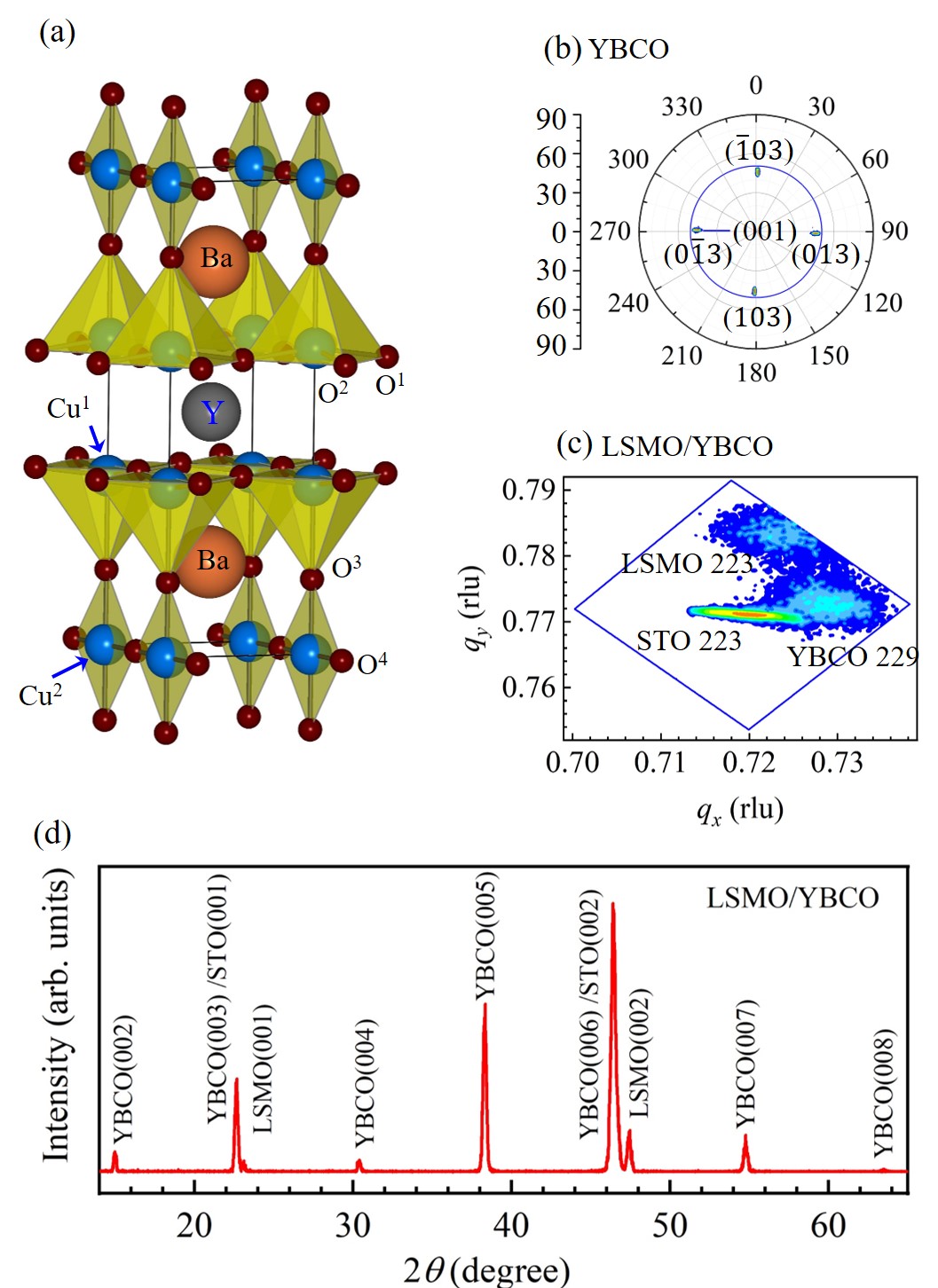}
%\vspace{-8ex}
\caption{(color online) (a) Crystal structure of YBa$_2$Cu$_3$O$_7$. (b) Pole figure plot of YBCO exhibiting growth along (001) direction. (c) Reciprocal space mapping of LSMO/YBCO show single crystallinity of the sample. (d) X-ray diffraction pattern of LSMO/YBCO film.}
\label{Fig1-str}
% \vspace{-2ex}
\end{figure}

High temperature superconductivity continues to be one of the most exciting field of research due to possibilities of realizing exotic physics and application in quantum technology. Among varied classes of superconductors, cuprates show the highest $T_c$ in ambient conditions and YBa$_2$Cu$_3$O$_x$ is the first material discovered with $T_c$ higher than 77 K \cite{wu, cava, gallagher}. YBCO forms in orthorhombic structure as shown in Fig. \ref{Fig1-str}(a). There are two non-equivalent Cu-sites (Cu$^1$ and Cu$^2$). Cu$^2$O$_4$ units are corner shared via O$^4$ sites to form chains. CuO planes are formed by Cu$^1$O$_5$ clusters corner shared via O$^1$ and O$^2$ (apical oxygen, O$^3$). These planes are separated by Y-layers and weakly hybridized leading to an effective two-dimensional electronic structure. Presence of both chains and planes of Cu-O networks makes this material highly complex relative to other cuprate superconductors. Majority of the Cu-O bond lengths are $\sim$~1.9 \AA\ except the Cu$^1$-O$^3$ ($\sim$ 2.3 \AA), a possible cleavage location \cite{cleaveexpt, cleavetheo}. YBCO ($x$ = 7.0) represents slightly overdoped case with $T_c$ close to 90 K. Hole concentration is normally reduced by removal of O$^4$ atoms which breaks Cu-O chains. Although YBCO believed to be a promising high $T_c$ candidate to probe the physics of unconventional superconductivity, it is studied relatively less compared to other cuprates due to several constraints such as complexity in tuning oxygen content and go beyond 7.0, preparation of a well-defined surface, etc.

Here, we explore the properties of YBCO thin film and it's heterostructure with La$_{0.67}$Sr$_{0.33}$MnO$_3$ (LSMO) on SrTiO$_3$ (STO) substrate. Such bilayer heterostructure of magnetic and superconducting materials are in focus for sometime to study the interplay of magnetism and superconductivity \cite{Zhang, chaudhuri2023interface, wisser2021growth, chen2009flux}, which are different from the crystalline materials having coexisting magnetic order and superconductivity \cite{pnictides}. LSMO and YBCO are good candidate materials for epitaxial growth with sharp interface due to their similar \textit{a, b} lattice parameters \cite{ankita-YBCO}. Various studies reported domain wall superconductivity \cite{houzet2006theory}, triplet superconductivity \cite{Samal}, vortex anomaly \cite{vortex}, exchange bias effect \cite{PrzyslupskiPRB}, additional flux pinning at the interface \cite{chaudhuri2023interface}, \emph{etc}. Electronic properties of the parent compounds, LSMO \cite{LSMO-Bindu} and YBCO \cite{YBCO} are sensitive function of charge carrier density. It is observed that tuning growth condition of YBCO, one can capture properties spanning the entire doping region of the phase diagram of crystalline YBCO \cite{Arpaia}. Here, we report direct evidence of hole transfer across the interface LSMO/YBCO bilayer.

%\section{Experimental details}

High quality single crystalline YBCO films (thickness $\sim$500 nm) were grown on STO(001) and STO(001)/LSMO(100 nm) surfaces using a homebuilt ultra high vacuum Pulsed Laser Deposition system \cite{ankita-YBCO}. Laser fluence, substrate temperature and oxygen pressure were set to 1.5 Jcm$^{-2}$, 750 $\degree$C, \& 300 mTorr, for LSMO and 1.8 Jcm$^{-2}$, 850 $\degree$C \& 400 mTorr for YBCO depositions, respectively. The films were \textit{in-situ} annealed for 1.5 hours under 500 Torr oxygen pressure in the deposition chamber and then slowly cooled down to room temperature. Pole figure data using a high resolution Panalytical X'Pert system exhibit excellent single crystallinity. For example, the pole figure for YBCO and reciprocal space mapping of LSMO/YBCO sample are shown in Fig. \ref{Fig1-str}(b) and (c), respectively. The quality of the films were further verified by x-ray diffraction (XRD) measurements as shown in Fig. \ref{Fig1-str}(d) for LSMO/YBCO sample exhibiting good crystallinity and growth along (00$l$) direction. Scanning electron microscopy data show smooth surface and the interface found to be sharp in the high resolution tunnelling electron microscopic data as given in Ref. \cite{ankita-YBCO}. Magnetization measurements show $T_c$ of about 89 K for YBCO film and 86 K for LSMO/YBCO bilayer \cite{ankita-YBCO}. Photoemission measurements were carried out at an energy resolution of 0.4 eV using a R4000 Scienta analyzer, monochromatic Al $K\alpha$ source and an open cycle helium cryostat from Advanced Research Systems, USA [base pressure $<$1$\times$10$^{-10}$ Torr].

%\section{Results and Discussion}

\begin{figure}
%\vspace{-2ex}
\centering
\includegraphics[width=0.45\textwidth]{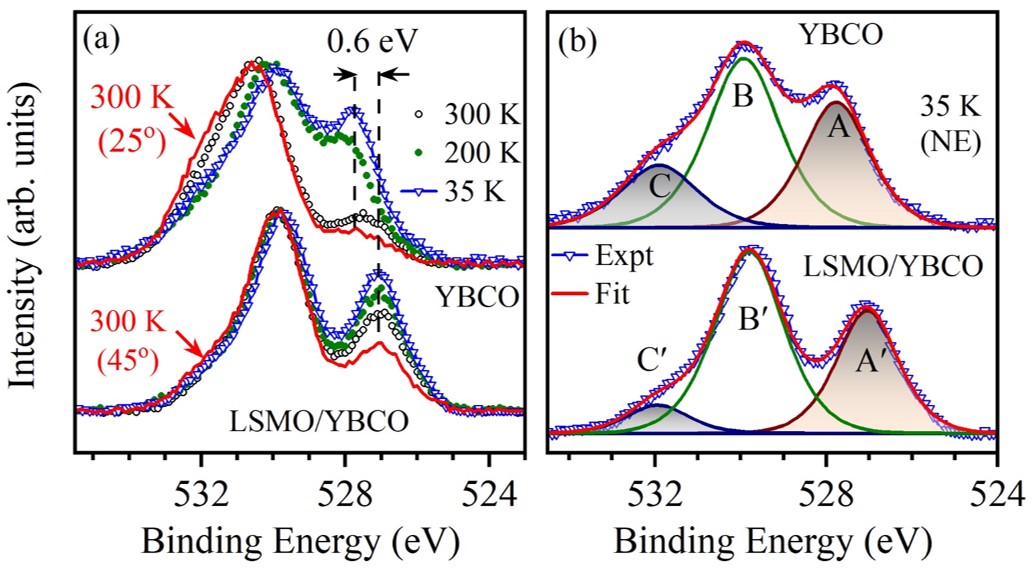}
%\vspace{-48ex}
\caption{(a) O 1$s$ spectra of YBCO (top panel) and LSMO/YBCO (bottom panel) at different temperatures and emission angles. (b) Simulation of the 35 K normal emission (NE) data of YBCO (top panel) and LSMO/YBCO (bottom panel).
}
\label{Fig2-O1s}
%\label{MH_2K}
\end{figure}

In Fig. \ref{Fig2-O1s}, we show the O 1$s$ core level spectra exhibiting three distinct features. The normal emission (NE) spectrum of YBCO at 300 K exhibit a feature, A at 527 eV and an intense broad feature, B around 530 eV. A change in emission angle to 25$^\circ$, thereby enhancing the surface sensitivity moderately \cite{surface}, leads to a decrease in intensity at 527 eV and an enhancement around 532 eV due to surface oxygens. Interestingly, 532 eV feature disappears at 200 K with significantly enhanced A; the data at 35 K exhibit further enhancement of A. The features in the LSMO/YBCO spectra shown in the lower panel in Fig. \ref{Fig2-O1s}(a) are more distinct and exhibit an energy shift of about 0.6 eV towards the Fermi level, $\varepsilon_F$. Change in emission angle to 45$^\circ$ leads to a reduced intensity at 526.4 eV along with a moderate increase around 531.5 eV. Cooling gradually enhances the intensity of the 526.4 eV feature.

We simulated the experimental data using the peaks, A, B, and C representing distinct features in the YBCO spectra (primed notations for LSMO/YBCO spectra). A typical case is shown in Fig. \ref{Fig2-O1s}(b) for the 35 K NE data in both the cases. The binding energy of the oxygen levels in YBCO have been studied extensively using hard x-ray photoemission \cite{YBCO-hike}, x-ray absorption \cite{O1s-XAS} and theory \cite{O1s-calcn}. In line with the reported results, the feature A is attributed to apical oxygens, O$^3$. Cu-O network formed by O$^1$ and O$^2$ in the Cu-O plane appear at higher binding energies. O$^4$ is least negative and has the highest binding energy. The feature B represents the O$^1$, O$^2$ and O$^4$ contributions. The peak positions observed here are somewhat different from the single crystal data  as the structural parameters of the films are influenced by the substrate. While the overall scenario is consistent with the literature, the gradual enhancement of A with cooling is puzzling and calls for further study.

\begin{figure}
\vspace{-2ex}
\centering
\includegraphics[width=0.45\textwidth]{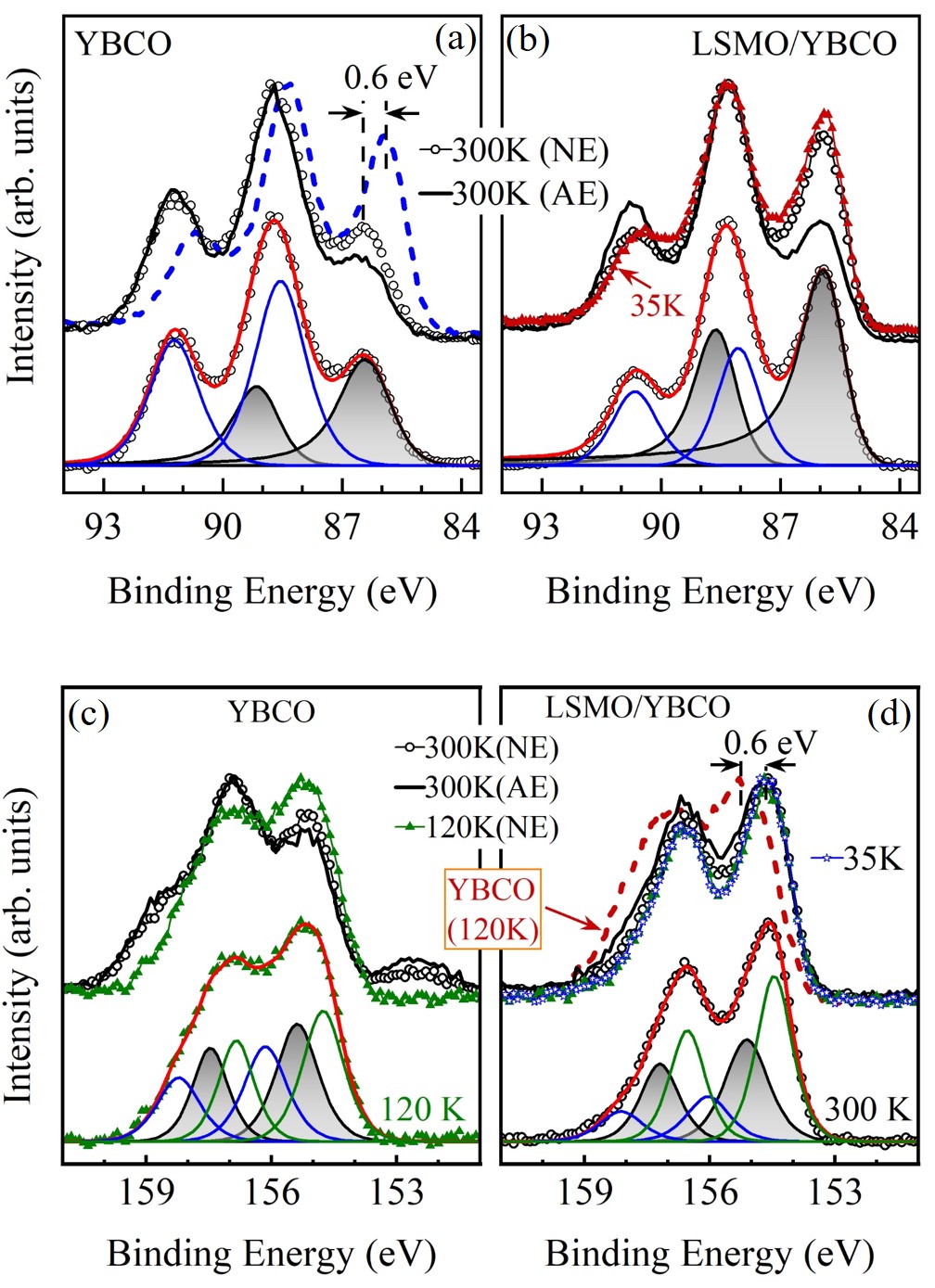}
%\vspace{-2ex}
\caption{Ba 4$d$ spectra of (a) YBCO and (b) LSMO/YBCO collected at normal emission (NE) and angled emission (AE). Dashed line in (a) is the LSMO/YBCO NE data. Lower panel show the fit of the 300 K NE spectra. Y 3$d$ spectra of (c) YBCO and (d) LSMO/YBCO at NE and AE. Dashed line in (d) is the YBCO 120 K NE data. The lower panels show the fit (120 K NE data of YBCO and 300 K NE data of LSMO/YBCO).}
\label{Fig3-Ba4dY3d}
\end{figure}

Ba 4$d$ and Y 3$d$ core level spectra are shown in Fig. \ref{Fig3-Ba4dY3d}. These elements are often termed as spectator elements \cite{spectator} as typically divalent Ba and trivalent Y have negligible contributions at $\varepsilon_F$. Ba 4$d$ data exhibit 3 distinct peaks in both the cases. The lower binding energy peak intensity is significantly smaller in the angled emission (AE) data attributing it to the bulk electronic structure; the surface peaks appear at higher binding energies \cite{ankita-YBCO}. We simulated the spectral functions using two sets of spin-orbit split features as shown in the lower panel of the figure for 300 K NE cases. In YBCO, the surface 4$d_{5/2}$ peak appears at 88.6 eV similar to divalent Ba \cite{XPS-BaO}. The bulk feature appears at 86.4 eV indicating a deviation from purely ionic behavior in the bulk \cite{spectator}. The spin-orbit splitting is found to be 2.8 eV. Ba 4$d$ data for LSMO/YBCO exhibit significantly intense bulk peak; for comparison, we superimposed the LSMO/YBCO 300 K NE data (dashed line) over the YBCO data in Fig. \ref{Fig3-Ba4dY3d}(a). Evidently, the electronic structure of YBCO got renormalized by the LSMO layer. In addition, the peak is shifted by about 0.6 eV towards $\varepsilon_F$ as shown in Fig. \ref{Fig3-Ba4dY3d}(a).

Y 3$d$ spectra shown in Fig. \ref{Fig3-Ba4dY3d}(c) and (d) exhibit additional features. YBCO spectra at 300 K show a peak at 152.5 eV which becomes stronger at angled emission \cite{ankita-YBCO}. It is absent in the 120 K data where the excess surface oxygen got desorbed. Clearly, Y spectra are most affected by the excess surface oxygens. Simulation of the Y 3$d$ spectra keeping the intensity ratio of spin-orbit split features fixed to their degeneracy requires three sets of spin-orbit split features. Typical fits are shown in the lower panel; Y 3$d_{5/2}$ peak in YBCO appears at 155 eV, 155.7 eV and 156.6 eV with spin-orbit splitting about 2.2 eV \cite{YBCO-hike, XPS-YBCO}. The peaks at 155 eV and 155.7 eV are the bulk and surface features. The peak at 156.6 eV is attributed to trivalent surface yttrium \cite{XPS-Y2O3}. The relative intensity of the spectral features in LSMO/YBCO are different from the YBCO data. The intensity of the bulk Y peak becomes stronger in the LSMO/YBCO sample similar to Ba case. A comparison of the YBCO and LSMO/YBCO spectra in Fig. \ref{Fig3-Ba4dY3d}(d) exhibits an energy shift of 0.6 eV. Similar energy shift in all the core level spectra suggests a shift of the Fermi level. Intensity of the bulk peak increases with cooling as also observed in O 1$s$ case.

\begin{figure}
\vspace{-2ex}
\centering
\includegraphics[width=0.45\textwidth]{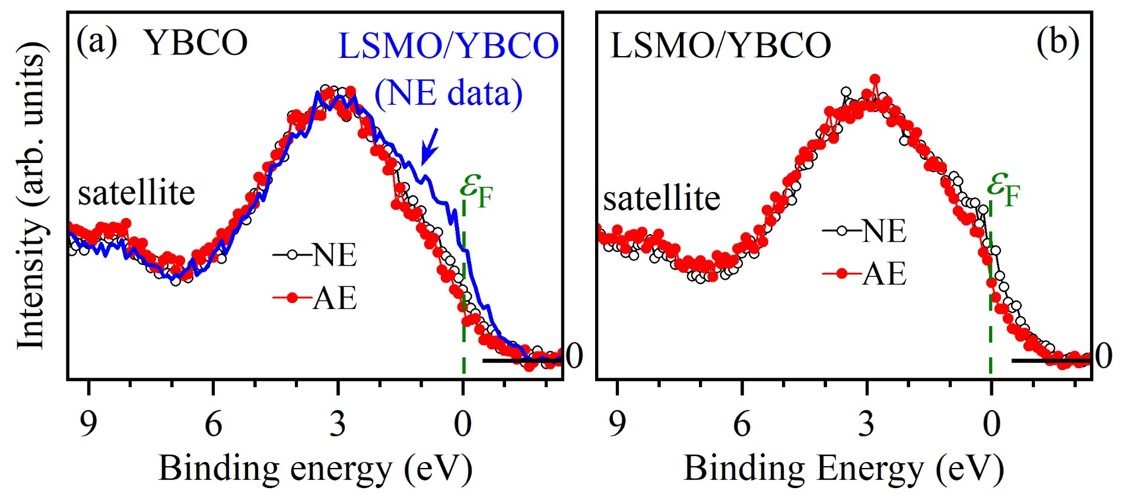}
%\vspace{-12ex}
\caption{(color online) Valence band spectra of (a) YBCO and (b) LSMO/YBCO heterostructure at room temperature. The line in (a) represents LSMO/YBCO NE data.
}
\label{Fig4-VB}
\end{figure}

The valence band spectra at different experimental conditions are shown in Fig. \ref{Fig4-VB} exhibiting distinct features around 8.2 eV, 3 eV and 1 eV binding energies. The feature at 8.2 eV is called satellite peak and corresponds to the poorly screened final state as the local potential will be larger than the well screened case. Presence of electron correlation will further modify the eigenenergy of such final states and often observed beyond 7 eV binding energies in cuprates \cite{Cu-chains-KBM}. The intensities below 6 eV binding energy represent the density of states (DOS) consisting of hybridized Cu 3$d$ - O 2$p$ states and non-bonding O 2$p$ states. Here, Cu 3$d$ contributions will be dominant as the photoemission cross section of the Cu 3$d$ states is significantly large at x-ray photon energies. In contrast to other cuprates, the valence band of YBCO has additional complexity due to the presence of both, Cu$^2$-O chains and Cu$^1$-O planes. A x-ray standing wave study \cite{XSW-PRB2015} show that the intensities near $\varepsilon_F$ are dominated by Cu$^1$ contributions. Cu$^2$ contributions appear essentially around 3 eV binding energies. Such descriptions are consistent with the valence band spectra of one-dimensional cuprates which are charge-transfer insulators and has no intensity at $\varepsilon_F$ \cite{Cu-chains-KBM}. The data at an emission angle of 25$\degree$ in YBCO exhibit diminished intensity near $\varepsilon_F$ reflecting surface-bulk differences in the electronic structure.

Comparison of the valence bands of YBCO and LSMO/YBCO at normal emission [see Fig. \ref{Fig4-VB}(a)] exhibits slightly weaker satellite feature at a binding energy lower than the YBCO case presumably due to Fermi level shift as observed in the core level spectra. The intensity at $\varepsilon_F$ is enhanced significantly in LSMO/YBCO sample. In YBCO, a hole-doped Mott insulator, $\varepsilon_F$ is pinned at the top of the valence band. Thus, an enhancement at $\varepsilon_F$ suggests large amount of hole transfer to YBCO across the interface \cite{ankita-YBCO, STO-ALO} which influences the electronic structure associated to Cu-O planes. Hole doping in YBCO is tuned via change in oxygen concentration, thereby, populating O$^4$ sites in the Cu$^2$-O chains. Hole-doping higher than that in  YBa$_2$Cu$_3$O$_7$ requires higher oxygen concentration which is difficult as all the oxygen vacancies in the Cu$^2$-O chains are filled. The energy shift of the core level spectra towards $\varepsilon_F$ and enhancement of spectral intensity at $\varepsilon_F$ indicates LSMO/YBCO heterostructure to be an unique case of overdoped YBCO keeping Cu-O chains and Cu-O planes intact.

\begin{figure}
%\vspace{-2ex}
\centering
\includegraphics[width=0.45\textwidth]{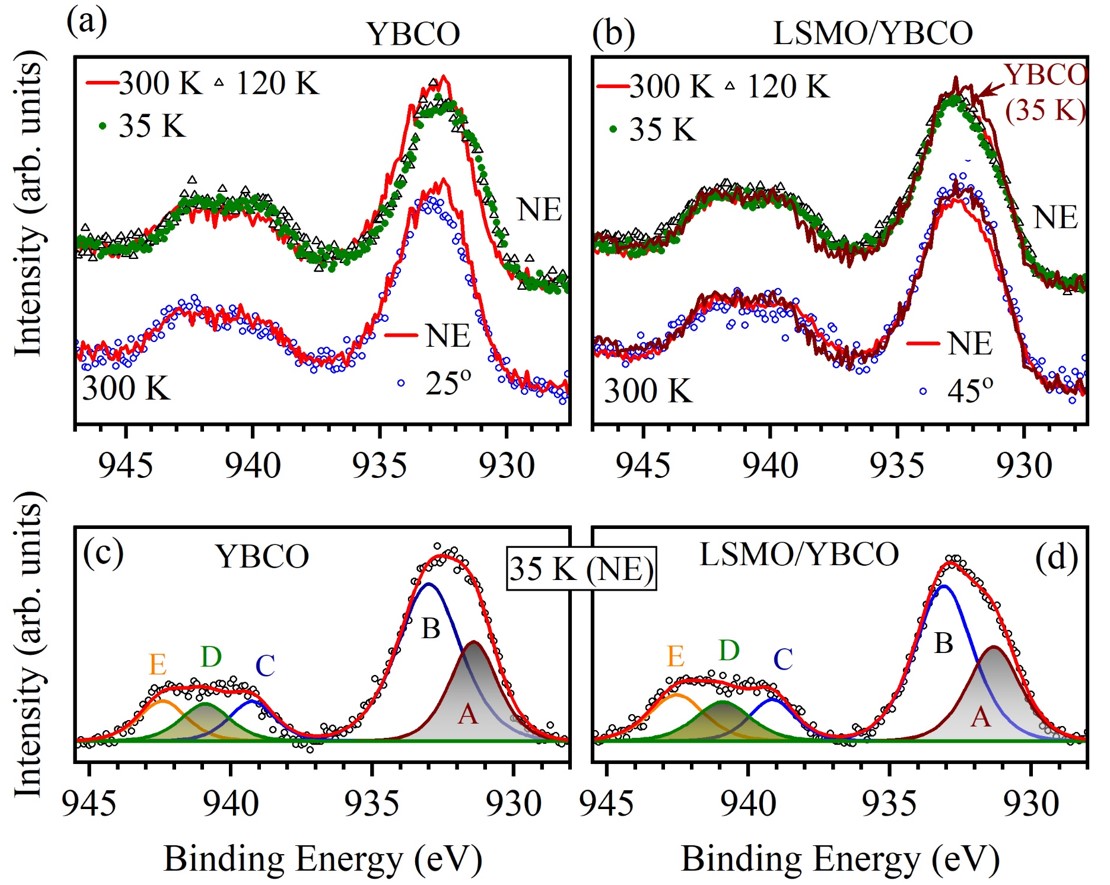}
%\vspace{-2ex}
\caption{Cu 2$p_{3/2}$ spectra of (a) YBCO and (b) LSMO/YBCO. Lower panel show the data collected at normal emission (NE) and angled emission geometry. YBCO spectrum at 35 K is superimposed over the LSMO/YBCO data in both panels of (b). Fit of the 35 K normal emission data of (c) YBCO and (d) LSMO/YBCO.
}
\label{Fig5-Cu2p3b2}
\end{figure}

To probe this further, we investigate Cu 2$p$ spectra in Fig. \ref{Fig5-Cu2p3b2}. The satellite features at 937 - 945 eV correspond to the poorly screened final states, ($|\underbar{2p}3d^{9}>$) and appear very similar in all the cases. The main peaks at 930 - 936 eV represent the well screened final states, $|\underbar{2p}3d^{10}\underbar{L}>$, where the positive charge due to the Cu 2$p$ core hole, $\underbar{2p}$ is screened via hopping of an electron to Cu 3$d$ level creating a ligand hole, $\underbar{L}$. The local potential at the photoemission site is sensitive to the location of $\underbar{L}$ leading to multiple features \cite{Cu2p-Boeske}. For example, in Li$_2$CuO$_2$ with edge-shared CuO cluster, $\underbar{L}$ is localized as edge-sharing does not allow inter-cluster hopping. This leads to a sharp peak at 934 eV. On the other hand, corner-shared CuO chains in Sr$_2$CuO$_3$ \cite{Cu2p-Boeske, Cu-chains-KBM} exhibit two features corresponding to localized $\underbar{L}$ (934.8 eV) and delocalized $\underbar{L}$ (933 eV). In some cases, an additional feature observed at even lower binding energies if $\underbar{L}$ forms a singlet with a Cu 3$d$ hole (Zhang-Rice singlet) \cite{ZRS}.

We observe that the YBCO spectral lineshape does not change at low temperatures and similar to the spectra in YBCO bulk crystals \cite{YBCO-hike} except 300 K spectrum exhibiting a narrower main peak. Usually, Cu$^+$ in Cu$_2$O (3$d^{10}$ configuration) or Cu$^{3+}$ (3$d^8$ configuration) in $D_{4h}$ crystal field show narrow peak due to the completely filled valence band \cite{Cu-Corelevel}. Considering the peak position of 932.5 eV is similar to the case in NaCuO$_2$, this observation suggests enhanced Cu$^{3+}$ contributions due to excess surface oxygens at 300 K. Comparison of the NE and AE data in Fig. \ref{Fig5-Cu2p3b2}(a) exhibit weaker intensity around 932 eV in the AE case which supports this view. Spectral lineshape in LSMO/YBCO are almost the same at all the temperatures studied and exhibit reduced intensity around 932 eV relative to YBCO data [see Fig. \ref{Fig5-Cu2p3b2}(b)]. The 300 K NE and AE data shown in the lower panel exhibit enhanced intensity of the main peak at 45$^\circ$ emission and becomes identical to the NE YBCO data. This indicates that the layers near the surface in LSMO/YBCO heterostructure is similar to YBCO bulk.

The fact that the core-hole screening occurs via charge transfer between Cu 3$d$ and oxygen 2$p$ bands, and also thereby to the neighboring CuO$_4$-plaquettes, an analysis of the core level data provides an opportunity to determine parameter values for the model Hamiltonian of the system,
$H = \sum_i \epsilon_{p}n_{ip}
+ \sum_{\mu\sigma}(\epsilon_{d} - U_{dc}n_{c})n_{\mu\sigma}
+\sum_{i \mu\sigma}(t_{pd}^{i\mu}d_{\mu\sigma}^{\dag} p_{i\sigma}+h.c.)
+\sum_{ij\sigma}(t_{pp}^{ij}p_{j\sigma}^{\dag} p_{i\sigma}+h.c.)$,
where $\epsilon_d$, $\epsilon_p$ and $t_{pd\sigma}^{i\mu} = <d_{\mu\sigma}|H|p_{i\sigma}>$ are Cu 3$d$, O 2$p$ and hopping energies, respectively. $U_{dc}$ is the Coulomb interaction strength between Cu 2$p$ hole and 3$d$ electrons. The charge transfer energy, $\Delta = \epsilon(d^{10}\underbar{L}) - \epsilon(d^9) = \epsilon_{d} - \epsilon_{p}$. Due to $D_{4h}$ crystal field, the valence levels have 3$d_{x^2-y^2}$ symmetry. The hopping strength, $t$ between Cu 3$d$ and symmetry adapted oxygen 2$p$ states will be, $t = 2t_{pd}$.
Therefore, the energy separation between the main peak and the satellite,
$\Delta{E}=\sqrt{(\Delta-U_{dc}+2t_{pp})^2 + 4t^2}$.
The ratio of satellite and main peak intensities, $I_{r} = tan^2(\theta - \phi)$, where
$sin^2\theta = {1\over{2}}\left[{1+{{(\Delta+2t_{pp})}/{\sqrt{(\Delta+2t_{pp})^2 + 4t^2}}}}\right]$
and
$sin^2\phi = {1\over{2}}\left[{1+{{(\Delta+2t_{pp}-U_{dc})}/{\Delta{E}}}}\right]$.

To find the constituent features, we simulated 35 K NE spectra as shown in Figs. \ref{Fig5-Cu2p3b2}(c) and (d). The features, A (931.4 eV) and B (933.0 eV) constitute the main peak and correspond to delocalized and localized $\underbar{L}$ cases. The simulation of the satellites requires three peaks, C, D and E at 939.2, 940.9 and 942.4 eV, respectively. The energy separation between the center of mass of the main peak and satellites in YBCO and LSMO/YBCO are similar and consistent with the reported values of $\sim$8.5 eV for YBCO \cite{DD-cuprates}. Therefore, $\epsilon_d$, $\epsilon_p$, $U_{dc}$, $t_{pd}$ and $t_{pp}$ are expected to be similar in these two systems. Considering $t_{pp}$ = 0.5 eV, $t$ = 1.5 eV and $U_{dc}$ $\sim$ 6-8 eV, the value of $\Delta$ is found to be about 3-5 eV, which places YBCO in the charge-transfer regime of the Zaanen-Swatzky-Allen phase diagram \cite{ZSA-phase}.
The integrated area of A, B, C, D \& E are 1.10, 2.00, 0.42, 0.48 \& 0.56 in LSMO/YBCO and 1.05, 2.21, 0.40, 0.40 \& 0.42 for YBCO. Intensity ratio of A to B is smaller in YBCO (= 0.475) than in LSMO/YBCO (= 0.55), which suggests larger density of nonlocal ligand holes in LSMO/YBCO. In LSMO/YBCO, $I_{r}$ (= 0.47) is much larger than $I_{r}$ (= 0.38) in YBCO; 3$d^9$ contribution is larger than 3$d^{10}\underbar{L}$ in LSMO/YBCO. These results suggests that the charge transfer across the interface in LSMO/YBCO enhances Cu 3$d$ hole content of the Cu-O plane. Such \textit{overdoped} condition explains the enhancement of intensity at the Fermi level as well as reduction of $T_c$ \cite{ankita-YBCO}.

%\section{Conclusion}

In summary, we studied the electronic structure of high quality single crystalline films of SrTiO$_3$/YBa$_2$Cu$_3$O$_7$ and SrTiO$_3$/La$_{0.67}$Sr$_{0.33}$MnO$_3$/YBa$_2$Cu$_3$O$_7$. While the overall electronic structure of YBCO observed here is consistent with the results from YBCO in bulk form, spectral evolution with temperature is puzzling and calls for further study. Valence band and core level spectra reveal signature of enhancement of hole concentration in the Cu-O planes in the LSMO/YBCO sample. This is a unique case providing a pathway of achieving overdoped condition in YBa$_2$Cu$_3$O$_7$ keeping the composition and structural parameters intact.

%\section*{Author Contributions}
%Ankita Singh: Methodology, Data curation and analysis. Sawani Datta: Data analysis. Ram Prakash Pandeya: Methodology. Srinivas C. Kandukuri: Methodology. Mahesh Gokhale: Methodology. Kalobaran Maiti: Conceptualization, Methodology, Data curation and analysis, Writing - original draft, Funding, all resources, and overall supervision.

%\section*{Conflicts of interest}
%There are no conflicts to declare.

%\section{Acknowledgements}
Authors acknowledge the financial support from the Department of Atomic Energy (DAE), Govt. of India (Project Identification no. RTI4003, DAE OM no. 1303/2/2019/R\&D-II/DAE/2079 dated 11.02.2020). K. M. acknowledges financial support from BRNS, DAE, Govt. of India under the DAE-SRC-OI Award (grant no. 21/08/2015-BRNS/10977)

\end{document}